\documentstyle[10pt,epsfig]{article}
\begin{document}

\begin {center}
{\Large $f_0(1370)$}

\vskip 5mm
{D.V.~Bugg\footnote{email address: D.Bugg@rl.ac.uk}},   \\
{Queen Mary, University of London, London E1\,4NS, UK}
\end {center}

\begin{abstract}
A summary is given of the main sets of data requiring the existence
of $f_0(1370)$.
Crystal Barrel data on $\bar pp \to \eta \eta \pi ^0$ contain a visible
$f_0(1370)$ peak and require at least a $19\sigma$ contribution.
This alone is sufficient to demonstrate its existence.
More extensive data on $\bar pp \to 3\pi ^0$ at rest contain delicate
interferences which determine the mass and width independently in
$^1S_0$ and $^3P_1$ annihilation and agree within 5 MeV for both mass and
width.
The peak in $2\pi$ is at $1282 \pm 5$ MeV, but the rapid increase
in $4\pi$ phase space with mass displaces the $4\pi$ peak to 1360 MeV.
BES II data for $J/\Psi \to \phi \pi^+\pi ^-$ contain a visible
$f_0(1370) \to 2\pi$ signal $>8\sigma$.
In all cases, a resonant phase variation  is required.

\vspace{5mm}

\end{abstract}

The $f_0(1370)$ plays a vital role in the spectroscopy of light
$J^P = 0^+$ mesons.
Several authors have questioned its existence,
though these criticisms are based on fits to very limited sets of
data.
To answer those questions, the best available data
have been refitted critically.
Full details of the analysis are given in \cite {f01370}.
Here essential points are summarised.

\begin {figure}  [h]
\begin {center}
\vskip -18mm
\epsfig{file=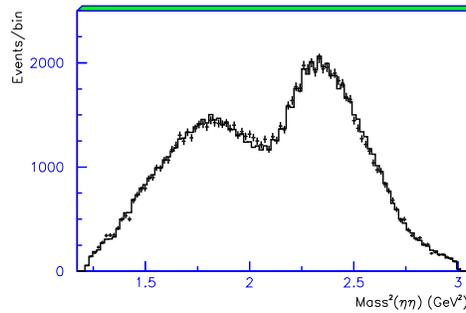,width=6.5cm}\
\vskip -6mm
\caption{The $\eta \eta$ mass projection for
$\bar pp \to \eta \eta \pi ^0$ at rest in liquid hydrogen}
\end {center}
\end {figure}

Crystal Barrel data on $\bar pp \to (\eta \eta)\pi ^0$ at rest
in liquid and gaseous hydrogen show two clear peaks in $\eta \eta$
at $\sim 1330$ and $1500$ MeV \cite {Amsler92}, see Fig. 1.
The low mass peak cannot be due to $f_2(1270)$, whose branching
ratio to $\eta \eta $ is very small.
A fit without $f_0(1370)$ is worse by 19 standard deviations, because
the peak at 1330 and the dip at 1430 MeV cannot be fitted by $\sigma
\to 4\pi$ and $a_0(980)$ alone.
These data alone are sufficient to demonstrate the existence of
$f_0(1370)$.

The data which determine resonance parameters best are Crystal
Barrel data on $\bar pp \to 3\pi ^0$ at rest in liquid and
gaseous hydrogen \cite {FurtherA}.
There is a conspicuous signal at low mass due to the $\sigma$
pole and high mass peaks due to $f_2(1270)$ and $f_0(1500)$.
The $f_0(1370)$ hides beneath the $f_2(1270)$,
but is clearly separated by angular analysis.
Interferences between the three $\pi \pi$ combinations determine
the mass and width of $f_0(1370)$ in a very delicate way.
The two sets of data allow a clean separation of annihilation
from $^1S_0$ and $^3P_1$ initial states.
The $f_0(1370)$ is at least a 32 standard deviation signal
in $^3S_1$ and 33 standard deviations in $^3P_1$.

The opening of the $4\pi$ channel plays an important
role.
The phase space for $\rho \rho$ and $\sigma \sigma$ are shown in
Fig. 2.
Relative contributions are poorly known,
so the fit to data finds the best compromise:
a Fermi function going to 1 at high mass.
Half-height is at $\sim 1.8$ GeV, and $4\pi$ inelasticity is
small at 1.3 GeV, ultimately the best mass.

\begin {figure}  [htb]
\begin {center}
\vskip -38mm
\epsfig{file=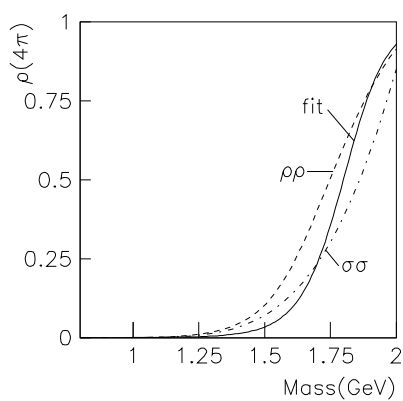,width=7cm}\
\vskip -10mm
\caption{$4\pi$ phase space for $\rho \rho$ (dashed),
$\sigma \sigma$ (chain curve) and the fit adopted here
(full curve)}
\end {center}
\end {figure}

The full form of the Breit-Wigner resonance formula,
\begin {equation}
f = 1/[M^2 - s - m(s) -iM\Gamma _{total}(s)]
\end {equation}
contains a real dispersive term $m(s)$ \cite {Nana}, which
for the $4\pi$ channel reads
\begin {equation}
m(s) =\frac {s - M^2}{\pi}\int \frac {ds'  M\Gamma
_{4\pi}(s')} {(s'-s)(s'-M^2)}.
\end {equation}
The slope of $m(s)$ near resonance is larger than
$(M^2 - s)$.
This point was not realised in earlier work.
Nevertheless, a good solution emerges naturally.
Loop diagrams for production of the $4\pi$ system behave like
vacuum polarisation and lead to strong renormalisation effects.
In consequence, only the ratio of $2\pi$ and $4\pi$ widths
is well determined: the absolute values can be varied through
a wide range, leaving the line-shape almost unchanged.

\begin {figure}  [htb]
\begin {center}
\vskip -14mm
\epsfig{file=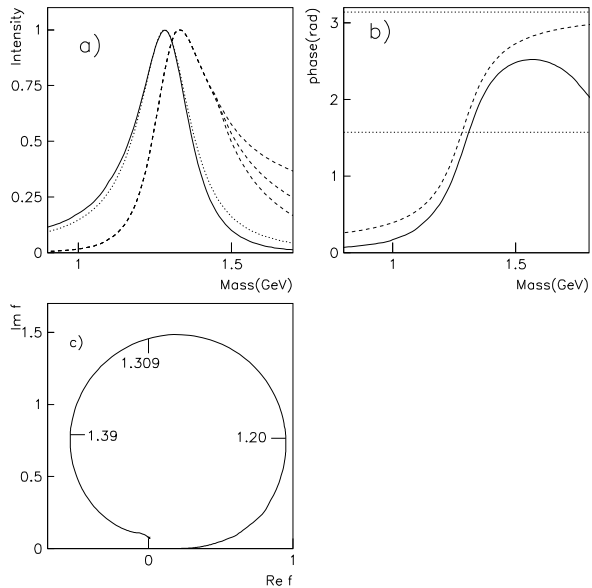,width=9cm}\
\vskip -8mm
\caption{(a) line-shapes of $f_0(1370)$ for $2\pi$ (full curve), a
Breit-Wigner amplitude with constant width (dotted), and for $4\pi$
(dashed);
(b) the phase angle measured from the bottom of the Argand plot
(full curve) and for a Breit-Wigner amplitude of constant width
(dashed); horizontal lines mark phase shifts of $\pi/2$ and
$\pi$,
(c) Argand plot; masses are shown in GeV.}
\end {center}
\end {figure}
Fig. 3 shows the essential points. The $\pi \pi$ peak is at
$1282 \pm 5$ MeV and is cut off towards higher massese by the
opening of the $4\pi$ channel.
The Breit-Wigner denominator is the same for $4\pi$ data, but
$4\pi$ phase space displaces the $4\pi$ peak upwards
by $78 \pm 10$ MeV.
This explains confusion in the Particle Data tables, where the
mass is quoted as 1200--1500 MeV.
One must distinguish between experiments fitting
(a) two-body channels and (b) $4\pi$ data.
The centre of the $4\pi$ peak at
half-height is at 1390 MeV, in close agreement with extensive Crystal
Barrel fits \cite {Thoma}.
The three dashed curves above 1500 MeV illustrate uncertainty in
$\Gamma (4\pi)$.

A remarkable feature of the new analysis is shown in Fig. 3(c).
Despite the strong dispersive effect of the $4\pi$ channel,
the amplitude follows a circle very closely.
The left-hand side of the loop is suppressed by coupling to $4\pi$.
The phase shift goes through $90^\circ$ at $1309 \pm 5$ MeV,
but the circle can be reproduced well with an effective mass
of 1282 MeV and a constant width of $207 \pm 15$ MeV.

\begin {figure}  [htb]
\begin {center}
\vskip -21mm
\epsfig{file=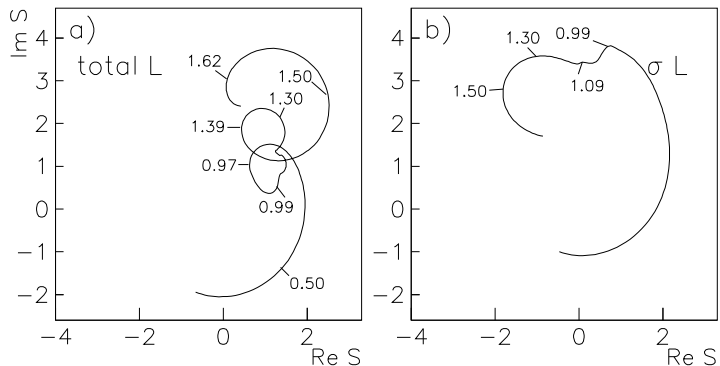,width=11cm}\
\vskip -54mm
\caption{Argand diagrams for the total $\pi \pi$ S-wave in
liquid hydrogen and $\sigma$ alone; masses are marked in GeV}
\end {center}
\end {figure}
Fig. 4 displays the Argand diagram for the $\pi \pi$ S-wave
in $^1S_0~\bar pp \to 3\pi ^0$.
There are successive loops due to the $\sigma$ pole, $f_0(980)$,
$f_0(1370)$ and $f_0(1500)$.
The third loop is the crucial one identifying $f_0(1370)$, or
$f_0(1300)$ to give it a new and improved mass.
An important test is to fit 40 MeV bins from 1100 to 1460 MeV
freely in magnitude and phase.
Real and imaginary parts of the amplitude move from the fitted
curve only by $\sim 15\%$ of the radius of the loop, consistent
with experimental errors.
This shows that the loop is definitely required.

A vital point in the new analysis is the inclusion of
$\sigma \to 4\pi$.
This cannot account for the $f_0(1370)$ loop, as illustated
on Fig. 4(b).
There is a loop near 1500 MeV due to this process, but it is
higher and much wider than $f_0(1370)$.
A weakness of all current fits to $4\pi$ data
is the omission of the $\sigma \to 4\pi$ amplitude.

\begin {figure}  [h]
\begin {center}
\vskip -15mm
\epsfig{file=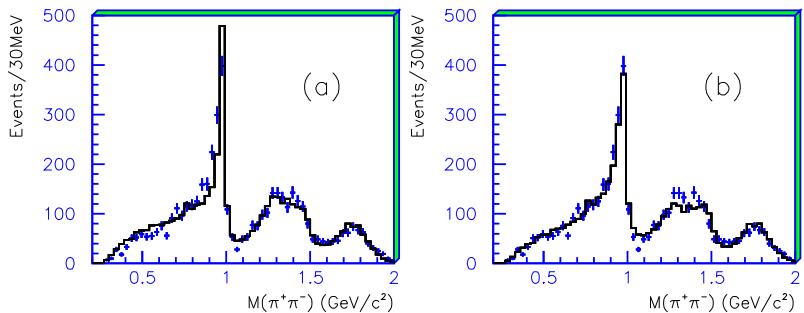,width=11cm}\
\vskip -6mm
\caption{Fits to BES II data on $J/\Psi \to \phi \pi ^+\pi ^-$.
(a) optimum fit;
(b) without $f_0(1370)$.}
\end {center}
\end {figure}
The $f_0(1370)$ and $f_0(1500)$ combine to produce a visible
peak in BES data for $J/\Psi \to \phi \pi ^+\pi ^-$
\cite {phipp}.
That  publication fitted the mass and width freely.
These data have now been  refitted using parameters fitted to
the $3\pi ^0$ and $\eta \eta \pi ^0$ data.
The fit of Fig. 5(a) is acceptable.
If the $f_0(1370)$ is removed, Fig. 5(b) shows the fit is
visibly poor.
Incidentally, the $f_0(1370)$ is well separated by angular
analysis from $f_2(1270)$ in these data; the latter optimises
with mass and width consistent with PDG values.

All the three sets of data discussed so far require a resonant
phase variation for $f_0(1370)$.
If the resonance is replaced with a peak of the same line-shape
but no phase variation (despite the fact that this is
non-analytic), $\chi^2$ is significantly worse in every case.

The $f_0(1370)$ also appears in GAMS data for $\pi ^+ \pi ^-
\to \pi ^0 \pi ^0$ at large $|t|$ \cite {Gams}
and in central production of $\pi\pi$ with parameters close
to those found here \cite {Barberis}.
Historically, it was first identified as $\epsilon (1300)$
in data from the Argonne and Brookhaven labs on $\pi \pi \to KK$
\cite {ANL}.
That identification was not clear-cut because
parameters of $f_0(980)$ and $\sigma \to KK$ were not known
accurately at that time.
Using modern values for those parameters, these early data
are entirely consistent with those fitted here \cite {Recon}.

The new fit includes the BES data for $J/\Psi \to \omega \pi \pi$
\cite {WPP} as an important constraint on the line-shape of the
$\sigma$ up to 1050 MeV; it is clearly visible by eye in the
$\pi \pi$ mass projection in those data.
The $\pi \pi$ phase shifts predicted by Caprini et al.
\cite {CCL}
using the Roy equations are also included.
The moments for Cern-Munich data on $\pi \pi \to \pi \pi$
\cite {Ochs} are also refitted.
Up to the $KK$ threshold, these data determine $\pi \pi$ phases
with errors of $\sim 3.5^\circ$.
However, above the $KK$ threshold, real and imaginary parts
of the amplitude become very strongly correlated
because differential cross sections alone do not separate
magnitude and phase.
The fit requires inclusion of some mixing between $\sigma$,
$f_0(1370)$ and $f_0(1500)$.
The Argand diagram for the $\pi \pi $ $I = 0$ S-wave is shown
in Fig. 6.
There is a loop due to $f_0(1370)$, crossing the Argand plot
rapidly at 1230 MeV.
The fit to Cern-Munich data gives $\Gamma_{2\pi}[f_0(1500)] =
61 \pm 5$ MeV.
The following branching ratios are also determined:
$\Gamma (f_0(1370) \to \eta \eta)/\Gamma (f_0(1370) \to \pi \pi )
= 0.19 \pm 0.07$,
$\Gamma (f_0(1500) \to \eta \eta)/\Gamma (f_0(1500) \to \pi \pi )
= 0.135 \pm 0.05$,
$\Gamma (\sigma  \to \eta \eta)/\Gamma (\sigma \to \pi \pi )
= 0.19 \pm 0.07$.

\begin {figure}  [htb]
\begin {center}
\vskip -16mm
\epsfig{file=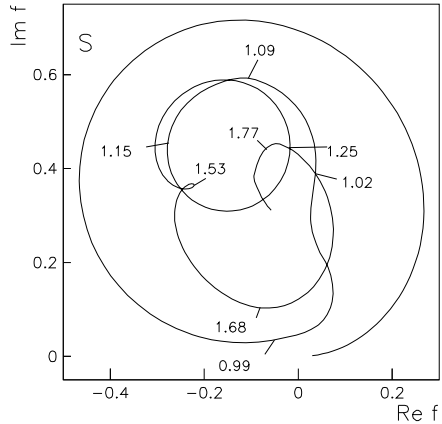,width=7cm}\
\vskip -6mm
\caption{Argand diagram for $\pi \pi$ $I=0$ S-wave in elastic
scattering}
\end {center}
\end {figure}

In summary, the $f_0(1370)$ is definitely required and
can be approximated for most purposes with a Breit-Wigner
denominator with $M = 1282 \pm 5$ MeV, $\Gamma = 207 \pm 15$
MeV, and appropriate phase space in the numerator for each
channel.
Together with $a_0(1450)$, $K_0(1430)$, $f_0(1710)$ and
$f_0(1500)$, one can complete a nonet together with the
$0^+$ glueball which mixes with the $q\bar q$ states.
The $f_0(1790)$ observed in BES data for $J/\Psi \to \phi \pi \pi$
(and several other sets of data) makes the first member of the
next nonet.
The exotic $\phi \omega$ peak observed by BES in $J/\Psi \to
\gamma \phi \omega$ \cite {omegaphi} is consistent with  the
upper half of $f_0(1790)$, suggesting it is locked to this
threshold at 1801 MeV, similar to the way $f_2(1565)$ is
locked to the $\omega \omega $ threshold \cite {Baker}.

\end {document}